\documentclass[journal]{IEEEtran}
\usepackage[pdftex]{graphicx}
\usepackage{optidef}
\usepackage{amsmath}
\usepackage{amstext}
\usepackage{algorithm}
\usepackage{algpseudocode}
\usepackage[utf8]{inputenc}
\usepackage[english]{babel}
\usepackage{amsthm}
\usepackage{amssymb}
\usepackage{xcolor}
\newtheorem{prop}{Proposition}

\usepackage{color}
\usepackage{bbding}
\DeclareMathOperator*{\argmax}{argmax}
\usepackage{makecell}
\usepackage{subfigure}

\begin{document}
\title{Maximizing Uplink and Downlink Transmissions in Wirelessly Powered IoT Networks}
%
\author{Xiaoyu Song and Kwan-Wu Chin
\thanks{Author Song and Chin are with the School of Electrical, Computer and Telecommunications Engineering, University of Wollongong. Emails: song-xiaoyu@outlook.com and kwanwu@uow.edu.au.}}
\maketitle

%
%
\begin{abstract}
This paper considers the problem of scheduling uplinks and downlinks transmissions in an Internet of Things (IoT) network that uses a mode-based time structure and Rate Splitting Multiple Access (RSMA).  Further, devices employ power splitting to harvest energy and receive data simultaneously from a Hybrid Access Point (HAP).
%
%
To this end, this paper outlines a Mixed Integer Linear Program (MILP) that can be employed by a HAP to optimize the following quantities over a given time horizon: (i) mode (downlink or uplink) of time slots, (ii) transmit power of each packet, (iii) power splitting ratio of devices, and (iv) decoding order in uplink slots.  The MILP yields the optimal number of packet transmissions over a given planning horizon given non-causal channel state information.  
We also present a learning based approach to determine the mode of each time slot using causal channel state information. 
The results show that the learning based approach achieves $90\%$ of the optimal number of packet transmissions, and the HAP receives $25\%$ more packets as compared to competing approaches. 
\end{abstract}
%

%
%

\section{Introduction}\label{sec:Introduction}
Internet of Things (IoT) networks have a broad range of applications, where they deploy a large number of devices to collect data from their environment such as temperature or the speed of cars. These data are then uploaded and analyzed by a server or base station. 
Further, these devices may be used to train machine learning models~\cite{9141214}.
Apart from that, some devices may have an actuator, where they are called upon by a controller to effect an environment after sensing some events~\cite{9860057}.
%
%
%

\begin{figure}[htbp]   
\centering   
\includegraphics[width=0.8\linewidth]{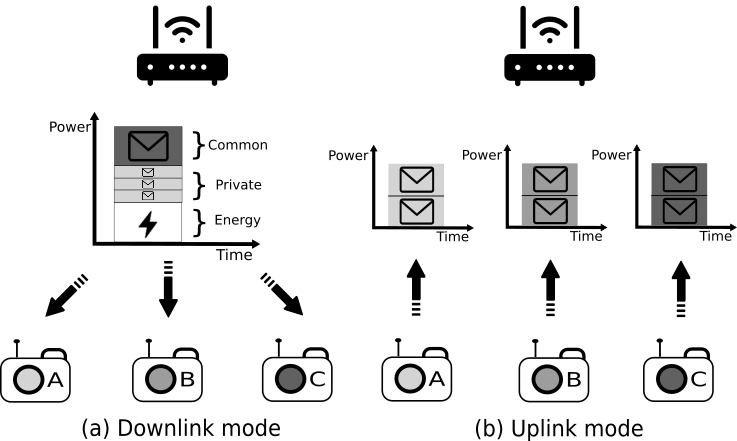}   
\caption{A HAP and $N$ RF-energy harvesting sensor devices. (a) In {\em downlink} mode, the HAP transmits data to devices as per RSMA. Devices simultaneously decode information and harvest energy using power splitting. (b) In {\em uplink} mode, devices transmit data to the HAP as per RSMA simultaneously.}   
\label{fig:system}   
\end{figure}

%
A fundamental issue in IoT networks is that devices have limited energy supply.  One solution is to deliver energy wirelessly. Specifically, devices harvest energy from radio frequency (RF) signals broadcasted by dedicated energy sources such as power beacons. For example, Powercast demonstrates that their P2100B receiver \cite{powercast} can harvest $2.75$ mW of power from RF signals. Further, since RF signals can carry both data and energy, the simultaneously wireless information and power transfer (SWIPT) architecture has received considerable attention, see \cite{8214104} for a survey. Briefly, SWIPT allows a device to harvest RF energy via time switching or power splitting. For time switching, a receiver switches between an energy harvesting circuit and an information decoding circuit. For power splitting, the received power is split into two parts, where one part is used for energy harvesting and the other for decoding data. 
Another issue is spectrum efficiency, where communications to/from devices may be limited by interference or multiple access schemes. 
To this end, rate splitting multiple access (RSMA) \cite{9831440} is now of interest. Briefly, RSMA divides a message into a {\em common} message and a {\em private} message. The common message contains information for all devices where each private message is for a specific device. Each device first decodes the common message and removes it from its received signal via successive interference cancellation (SIC)~\cite{1421925}. After that, a device decodes its private message by treating all other private messages as noise. Similarly, RSMA can also be applied for uplink transmissions \cite{485709}. Specifically, a device divides its uplink message into two sub-messages and transmits them to a base station with a different power level.  The base station then decodes these messages using SIC.  
To date, no works have considered uplink and downlink transmissions in an RF-powered RSMA IoT network.
Fig.~\ref{fig:system} shows an example network. A hybrid access point (HAP) transfers energy and exchanges information with a set of sensor devices. Unlike prior works, see Section~\ref{sec:RelatedWorks}, we do not consider the conventional time-division duplex (TDD) structure. Instead, time slots are used either for {\em downlinks} or {\em uplinks}. Specifically, when the HAP marks a time slot as downlink, it transmits data to devices as per RSMA. Further, devices adjust their power split ratio to decode information and harvest energy simultaneously. As for time slots marked as uplink mode, devices consume their harvested energy to transmit data packets to the HAP using RSMA.
Note that the said time mode structure allows an HAP to determine whether a time slot is better for downlinks or uplinks; i.e., if in time slot $t$ the channel gain from devices is poor but the channel gain to devices is good, then time slot $t$ can be used to transfer more energy and data to devices, and vice-versa.  This means the HAP is able to defer uplinks to future time slots that have better channel gains, and thus allow devices to use their precious harvested energy efficiently.

\begin{figure}[htbp]      
\centering   
\includegraphics[width=0.8\linewidth]{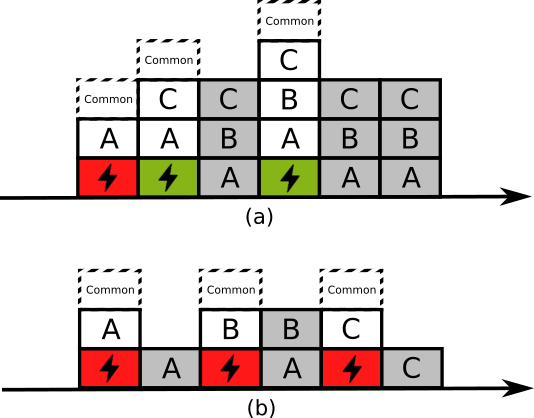}   
\caption{Two examples of charging and data collection schedule: (a) mode selection structure, and (b) conventional TDD frame structure.}   
\label{fig:schedule}   
\end{figure}

Our aim is to maximize the number of packets decoded by the HAP and devices over a given planning horizon with {\em multiple} time slots. To do this, the HAP needs to decide the mode of each slot. Specifically, it needs to determine whether a slot is better suited for downlink or uplink transmissions.  
%
Fig.~\ref{fig:schedule} shows a possible schedule using a mode based time structure versus a conventional TDD frame. Assume the energy delivery in red colored slots is over a poor channel; otherwise, it is colored in green. Each white colored block indicates that a packet is transmitted and successfully decoded whereas a dotted block means a common message is transmitted in the downlink. We see that the channel condition in the first slot is poor for energy delivery which results in a low amount of harvested energy. Consequently, for the TDD structure, only one uplink packet is transmitted in the second slot. By contrast, in our mode selection structure, the HAP selects downlink mode in the second slot, which has a good channel condition. This thus allows devices to harvest more energy and transmit three packets in the third slot. Furthermore, the HAP selects downlink mode in the fourth slot because the channel is good for downlink transmissions, which results in the transmission of four packets. In contrast, the HAP transmits two packets in the third slot if it uses the TDD structure. As a result, we see that our mode selection structure has nine more packets than the conventional TDD structure.

Henceforth, we aim to design a scheduler that is run by a HAP. The scheduler determines (i) the mode of each slot, (ii) the transmit power of each packet, (iii) the power splitting ratio at each device in a downlink slot, and (iv) the decoding order used by the HAP in uplink slots.
Designing such a scheduler is challenging as it has to consider RSMA and RF charging jointly.  
In fact, the said combination introduces a fundamental coupling relating to the HAP's RF charging process, which affects downlinks and RF energy harvested by devices, which in turn affects uplinks.
To elaborate, (i) a HAP must consider the fact that the energy at devices is coupled across time slots. Specifically, the energy at devices is a function of the channel gain from the HAP, numbers of slots used for energy delivery and the data transmissions in previous time slots, (ii) a HAP does not have future channel state information, meaning it does not know whether there is a future time slot that is better for downlink or/and uplink transmissions, and (iii) the transmit power of each split message in both downlink and uplink slots has to satisfy a given SINR condition. Critically, the transmit power will be affected by the set of transmitting devices and the decoding order of received packets.
%

To this end, we make the following contributions:
\begin{itemize}
%
\item We are the first to study a {\em novel} RSMA-aided SWIPT wireless powered IoT network.   
In particular, to the best of our knowledge, no prior works have considered the aforementioned issues and challenges that arise from the combination of RSMA and RF-charging in order to maximize the number of uplink and downlink packets in the said network.
%
%
%
\item We outline the {\em first} mixed integer linear program (MILP) that can be used to compute (i) the mode of each slot over a planning horizon, (ii) the transmit power allocated for each packet, (iii) the power splitting ratio of each device in downlink slots and (iv) the decoding order used by the HAP in uplink slots. Advantageously, it yields the optimal number of messages received by devices and the HAP given non-causal channel state information.
\item We show how a novel reinforcement learning based approach can be used to determine the mode of each time slot over a planning horizon.  Another key innovation is that for each downlink and uplink slot, it uses linear programs to determine the transmit power of each packet, the power splitting ratio at devices and the decoding order of uplink packets.  
\item We present the first study of the said solutions.  The results show that the reinforcement learning approach reaches $90\%$ of the optimal number of packet transmissions of the said MILP. Further, reinforcement learning leads to $15\%$ and $25\%$ higher number of packet transmissions than a round robin  and random approach, respectively.
\end{itemize}
\begin{table*}[htbp]  
\centering  
\caption{A comparison of related works.}  
\label{tab:references}  
\begin{tabular}{|l|c|c|c|c|c|}  
\hline {\bf References} & \makecell{{\bf RF Charging} \\ {\bf (Power splitting)} }& {\bf Multi-slot} & \makecell{{\bf Joint uplink} \\ {\bf and downlink}}  & {\bf Mode} & {\bf RSMA}  \\ 
\hline \hline \cite{9528052,9248582, s11276-019-02126-z, 9627180} & \Checkmark (\Checkmark) & \XSolid & \XSolid & \XSolid & \Checkmark \\
\hline \cite{8491048, 8815494, 9195473} & \Checkmark (\XSolid) & \XSolid & \XSolid & \XSolid & \Checkmark \\
\hline \cite{7555358, 9201435, 9461768, 9663192, 9676684, 9257190} & \XSolid (\XSolid) & \XSolid & \XSolid & \XSolid & \Checkmark \\
\hline \cite{9562976} & \XSolid (\XSolid) & \Checkmark & \XSolid & \XSolid & \Checkmark \\
\hline \makecell[l]{\cite{7956255, 6884177, 7833146, 8854318, 7874074,8472716}, \\ \cite{8294215, 8695878, S1874490721000616, 7982605}} & \Checkmark (\Checkmark) & \XSolid & \Checkmark & \XSolid & \XSolid \\
\hline \makecell[l]{\cite{9154671, 8855977, 8947043, 8536389, 8292385,8650384}, \\ \cite{ S1389128620311476,7848950}} & \Checkmark (\XSolid) & \XSolid & \XSolid & \XSolid & \XSolid \\
\hline \cite{7990524} & \Checkmark (\XSolid) & \Checkmark & \Checkmark & \XSolid & \XSolid \\
\hline \cite{10109156,9367293} & \Checkmark (\XSolid) & \Checkmark & \XSolid & \Checkmark & \XSolid \\
\hline {\bf Our work} & \Checkmark (\Checkmark)& \Checkmark & \Checkmark & \Checkmark & \Checkmark \\ \hline 
\end{tabular}  
\end{table*} 
Next, Section~\ref{sec:RelatedWorks} discusses prior works.  We then formalize the system and problem in Section~\ref{sec:SysM} and \ref{sec:MILP}, respectively.  We then present a learning solution in \ref{sec:Q-Learning}.  Our evaluation of the said MILP and learning solution is detailed in \ref{sec:Eval}.   Lastly, we conclude in Section~\ref{sec:conclusion}.

%
\section{Related works}\label{sec:RelatedWorks}
Our research overlaps with works  (\romannumeral1) that consider joint uplink and downlink optimization in wireless powered communication networks (WPCNs), and (\romannumeral2) data transmissions using RSMA. 

\subsection{Joint Uplink and Downlink}
Most works consider a Time-Division Duplex (TDD) frame structure where a downlink phase must be followed by an uplink phase.  
%
In \cite{7956255}, the authors aim to optimize the beamforming weight at HAPs and devices, power splitting ratio at each device, individual data rate of each user and downlink duration in order to maximize the uplink sum-rate of the system. The authors in \cite{7874074} aim to maximize uplink and downlink sum-rate by optimizing the beamformer at an HAP and devices. In \cite{ S1874490721000616, 7982605}, the proposed solution optimizes the time duration for uplink and downlink and power split ratio at each device. 
Some works consider time switching, where an HAP uses different time slots to deliver energy and transmits data to devices. As an example, the solution in \cite{8292385} maximizes sum-rate by optimizing the time allocation for downlink and uplink. In \cite{8855977, 8947043}, the authors aim to maximize both uplink and downlink sum-rate by optimizing charging, downlink and uplink duration and transmit power of devices. 
The authors of \cite{S1389128620311476} consider clustering of devices; each cluster uses a different subcarrier. The authors maximize the system sum-rate by optimizing the subcarrier assigned to devices, time duration for energy and data delivery and transmit power of devices.

There is also research into orthogonal frequency division multiple access (OFDMA) networks. The authors of \cite{8536389} aim to maximize uplink sum-rate while ensuring a minimum downlink data rate. They achieve this by optimizing subcarrier assignments and transmit power of devices in downlink slots and the time duration of uplink slots. In \cite{8472716, 8388904}, the authors assign subcarriers and determine the transmit power over each sub-carrier and power-splitting ratio at devices.  Reference~\cite{8472716} considers uplink sum-rate whereas in \cite{ 8388904}, the authors consider both uplink and downlink transmissions.
Other than sum-rate, other metrics of interest include energy consumption, fairness among device, and energy efficiency. As an example, in \cite{ 7990524}, the authors aim to minimize the energy consumption of both uplink and downlink transmissions. They consider a HAP with multiple antennas and control its beamforming power and duration. The authors in \cite{7833146} also consider a TDD frame where they address the well-known double near far effect~\cite{6678102} by maximizing the minimum downlink and uplink sum-rate. The authors of \cite{8294215} aim to optimize the uplink and downlink duration, transmit power of the HAP and devices and the power splitting ratio in order to maximize fairness among devices. The authors in \cite{6884177} study uplink and downlink energy efficiency and aim to optimize device selection and transmit power at the HAP and devices.

\subsection{Rate Splitting Multiple Access}
Many works have studied resource allocation in RSMA networks.  However, none of them consider using RSMA for {\em both} downlink and uplink communications.  In fact, many previous works only consider RSMA for downlink transmissions.  In this respect, the transmitter in \cite{8815494,9528052} exploits beamforming and RSMA to transmit information and energy to devices. The transmitter in \cite{8815494} has perfect channel state information where in \cite{9528052}, the transmitter has imperfect channel state information; i.e., it only has historical or probability distribution of past channel gains. In \cite{9528052}, the aim is to maximize energy efficiency by optimizing beamforming and power splitting ratio. The authors in \cite{8815494} optimize beamforming at a transmitter to maximize its downlink sum rate and also ensure receivers harvest a minimum amount of energy. In \cite{9195473}, an HAP with multiple antennas operates in a multicell network, where it aims to maximize both energy and spectrum efficiency by jointly optimizing beamforming weight and the rate of common and private messages. In \cite{10198464}, the authors aim to improve the spectrum efficiency of a two-tier network by optimizing transmit precoding and rate allocation.
Some works study the theoretical maximum sum-rate or minimum transmit power at an HAP. For example, in \cite{8491048}, the authors analyze the sum-rate of an energy harvesting network where a relay uses its harvested energy to forward received signals using RSMA. 
The work in \cite{9248582, s11276-019-02126-z} minimizes the transmit power of a base station by finding the optimal transmit power, rate of messages and power splitting ratio. In \cite{9627180}, the authors also aim to find the minimum transmit power where they consider a full-duplex system. 

Many solutions are designed to maximize the downlink sum-rate of RSMA networks where devices do not harvest RF energy.  One such solution is~\cite{9562976}, where the authors assume a base station that has imperfect channel state information and aim to maximize long-term sum-rate by optimizing the transmit power of common and private messages in each time slot. In \cite{7555358}, the authors also maximize sum-rate using imperfect channel state information where they consider a base station with multiple antennas. The authors of \cite{9201435} study sum-rate maximization in a multicarrier system where they aim to optimize the transmit power allocated to subcarriers. The authors in \cite{9461768} first use RSMA under a SIC constraint and aim to optimize the rate and transmit power of private messages. The work in \cite{9663192} considers devices with multiple antennas, and aim to optimize the precoder of messages. 

Some works also consider uplink transmissions as per RSMA.  An example work is~\cite{9676684}, where the authors analyze the throughput and outage probability of a RSMA network with only two devices. In \cite{9257190}, the authors aim to maximize the uplink sum-rate at a HAP by optimizing the transmit power of devices and the decoding order at the HAP.

\subsection{Research Gaps}
To summarize, referring to Table~\ref{tab:references}, we see only reference \cite{9367293} and \cite{10109156} have considered a mode-based structure but they do not consider devices with power splitting capability.  A key distinction is that our work is the first to consider joint uplink and downlink optimization for a RSMA-based wireless powered IoT network that uses a mode-based time structure.  
Further, only a few works have studied optimization over multiple time slots. However, these works have a different system setup. As we see in Table~\ref{tab:references}, reference \cite{9562976} does not consider RF charging. In addition, the work in~\cite{9367293} and \cite{10109156} does not consider RSMA.

\begin{table}[htbp] 
\begin{center} 
\caption{Table of Notations.} 
\label{notation} 
\begin{tabular}[]{l l} 
\hline 
\bf Notation & \bf Description \\ \hline 
\textbf{1.} & \textbf{System model} \\ \hline
$\mathcal{N}$ & A set of devices \\ 
$n$ & Device index \\
$\mathcal{T}$ & A set of all time slots \\
$T,t$ & Time horizon and time slot index\\
$\tau$ & Slot duration \\
$\hat{I}^t,\breve{I}^t$ & Mode of a slot; {\em Uplink} or {\em downlink}  \\
$g^t_n$ & Channel gain \\
$\theta^t_n$ & Short-term fading \\
$\beta$ & Path-loss exponent \\
\hline
\textbf{2.} & \textbf{Downlink mode} \\ 
\hline
$P^t$ & HAP transmit power \\
$\mu^t_c,\mu^t_n$ & Power coefficient of a common and private message \\
$\rho^t_n$ & Power splitting ratio \\
$\gamma_{c,n}^t,\gamma_n^t$ & The SINR of a common and private message \\
$\psi()$ & Non-linear conversion model \\
$M$ & Maximum received power \\ 
$a,b$ & Energy harvesting circuit model parameters \\
$H_n^t$ & Energy harvested by device $n$ in slot $t$ \\
$\Gamma$ & SINR threshold \\
$C_n^t,D_n^t$ & \makecell[l]{A binary variable for common and private message of \\ user $n$ in slot $t$}\\
\hline
\textbf{3.} & \textbf{Uplink mode} \\ 
\hline
$p^t_{n,j}$ & Transmit power of the $j$-th message of device $n$ \\
$U_{n,j}^t$ & Indicator of the $i$-th message from devices $n$ in slot $t$\\
$\mathcal{K}$ & A set of SIC decoding orders \\
$O^t_k$ & Indicator of the $k$-th decoding order\\
$\pi^t_{n,j}$ & Position of message $s_{n,j}$ in decoding order $k$ \\
$\gamma_{n,j}^{t,k}$ & SINR of message $s_{n,j}$ in decoding order $k$\\
\hline
\textbf{4.} & \textbf{Energy and sum-rate} \\ 
\hline
$E_n^t$ & Residual energy of device $n$ in slot $t$ \\
$R$ & Number of transmissions over $T$ slots \\
\hline
\end{tabular} 
\end{center}
\end{table}  
%

%
%
\section{system model}\label{sec:SysM}
Table~\ref{notation} summarizes our notations. 
There is a HAP and a set $\mathcal{N} =\{1,2,\ldots, N\}$ of RF harvesting devices, indexed by $n$. The distance between device $n$ and the HAP is $d_{n}$.  Time is divided into slots, where each slot has index $t$ and a duration of $\tau$ second. Define $\mathcal{T} = \{1,2,\ldots, T\}$ to to be a planning horizon with $T$ time slots.   
The transmit power of the HAP in slot $t$ is $P^t$. The maximum transmit power of the HAP and each device is denoted as $P_0$ and $p_0$, respectively. 
We assume both devices and HAP have saturated traffic, meaning they always have a message to transmit.

Each slot can either be in Downlink or {\em Uplink} mode. Let $\breve{I}^t$ and $\hat{I}^t$ be binary variables, where $\breve{I}^t=1$ means slot $t$ is in Downlink mode and $\hat{I}^t=1$ indicates slot $t$ is in Uplink mode. The HAP selects one mode in each time slot, whereby $\breve{I}^t$ and $\hat{I}^t$ are constrained by
    \begin{equation} \label{mode constraint}
        \breve{I}^t+\hat{I}^t = 1, \quad \forall t \in \mathcal{T}.
\end{equation}
We assume block fading, where the channel power gain $g^t_{n}$ remains fixed during each time slot but varies across time slots. In slot $t$ the channel between device $n$ and the HAP is modeled as 
\begin{equation}\label{channel}
    g^t_{n} = \theta^t_n d^{-\beta}_{n},
\end{equation}
where $\theta^t_n$ represents short-term fading and it is assumed to be Exponentially distributed with unity mean, and $\beta$ is the path-loss exponent.
In a Downlink slot, the HAP transmits messages to devices using RSMA, where the message for device $n$ is denoted as $W_n$. Each message is split into a {\em common} part and a {\em private} part, denoted as $W^c_n$ and $W^p_n$, respectively. The HAP uses a public codebook to combine the common part of each user into a single common message, denoted as $W^c$. The common message $W^c$ and $N$ private messages are encoded into symbols independently, denoted as $s_c$ and $s_n$, respectively. The HAP then uses superposition coding to transmit a composite signal to all users. The composite signal transmitted by the HAP is denoted as
\begin{equation} \label{equ:transmitted signal}
    \mathbf{x} = \sqrt{\mu^t_c P^t} s_c+\sum_{n=1}^N \sqrt{\mu^t_n P^t} s_n,
\end{equation} 
where $\mu^t_c$ and $\mu^t_n$ denote the power coefficients of messages, and they are constrained by 
\begin{equation}\label{downlink power}
    \mu^t_c + \sum_{n=1}^N \mu^t_n  \le \breve{I}^t, \quad \forall t \in \mathcal{T}.
\end{equation} 
Thus, the signal received by device $n$ is denoted as
\begin{equation} \label{equ:received signal}
    \mathbf{x^{\prime}} = \sqrt{\mu^t_c P^tg^t_n} s_c+\sum_{n=1}^N \sqrt{\mu^t_n P^tg^t_n} s_n + N_k,
\end{equation} 
where $N_k$ is ambient noise in the channel. Note that the power coefficients are non-negative if slot $t$ is in Downlink mode.

To receive messages from the HAP, a device first decodes the common message $W^c$ from the received signal by treating all private messages as interference. The common message $W^c$ is successfully decoded by device $n$ if its SINR at device $n$, denoted as $\gamma_{c,n}^t$, satisfies
\begin{equation}\label{equ:gamma_cn}
    \gamma_{c,n}^t = \frac{\mu^t_c P^t g^t_n}{\sum_{n=1}^N \mu^t_n P^t g^t_n + N_o} \ge \Gamma,
\end{equation}
where $N_o$ is the noise power. Device $n$ then maps $W^c$ back to $W^c_n$ according to the said public codebook. After that, device $n$ cancels the common message $W^c$ from the received signal as per SIC \cite{1421925} and then decodes its private message $W^p_n$. The private message $W^p_n$ is successfully decoded by device $n$ if its SINR, denoted as $\gamma_{n}^t$, satisfies
\begin{equation}\label{equ:gamma_n}
    \gamma_{n}^t = \frac{\mu^t_n P^t g^t_n}{\sum _{m\neq n} \mu^t_m P^t g^t_m + N_o} \ge \Gamma.
\end{equation}
Let $C^t_n$ be a decision variable to indicate if device $n$ decodes a common message in slot $t$, where $C^t_n = 1$ means that a common message is successfully decoded by device $n$ in slot $t$. Let $D^t_n$ be a decision variable to indicate if device $n$ decodes a private message in slot $t$, where $D^t_n = 1$ means that a private message is successfully decoded by device $n$ in slot $t$. The previous facts are formalized as follows:
\begin{equation}\label{common SINR}
    \gamma_{c,n}^t + (1-C^t_n)\Phi \ge \Gamma, \quad \forall t \in \mathcal{T}, \forall n \in \mathcal{N},
\end{equation}
and
\begin{equation}\label{private SINR}
    \gamma_{n}^t + (1-D^t_n)\Phi \ge \Gamma, \quad \forall t \in \mathcal{T}, \forall n \in \mathcal{N}.
\end{equation}
In addition, we have
\begin{equation}\label{downlink transmission}
    C^t_n \le \breve{I}^t, \quad \forall t \in \mathcal{T}, \forall n \in \mathcal{N},    
\end{equation}
and
\begin{equation}\label{downlink transmission 1}
    D^t_n \le \breve{I}^t, \quad \forall t \in \mathcal{T}, \forall n \in \mathcal{N},
\end{equation}
which ensure $C^t_n$ and $D^t_n$ are set to zero when slot $t$ is not in Downlink mode. The number of downlink transmissions for device $n$ in slot $t$ is then calculated as $\breve{R}^t_n = C^t_n + D^t_n.$
Each device supports power splitting \cite{6489506} where a received signal is split into two parts. 
Define the power splitting ratio of device $n$ as $\rho^t_n\in [0,1]$.
A device $n$ uses $\rho^t_n$ of the received signal power for data transmission. This means the SINR of the common message $W^c$ is
\begin{equation}\label{common SINR 2}
    \gamma_{c,n}^t = \frac{\mu^t_c \rho^t_n P^t g^t_n}{\sum_{n=1}^N \mu^t_n \rho^t_nP^t g^t_n + N_o}.
\end{equation}
Similarly, the SINR of private message $W^p_n$ is 
\begin{equation}\label{private SINR 2}
    \gamma_{n}^t = \frac{\mu^t_n \rho^t_n P^t g^t_n}{\sum _{m\neq n} \mu^t_m \rho^t_nP^t g^t_m + N_o}.
\end{equation}
The power splitting ratio $\rho^t_n$ is allowed to be greater than zero when slot $t$ is in Downlink mode; otherwise, it is set to zero.
Formally, 
\begin{equation} \label{PS constraint}
    \rho^t_n \le \breve{I}^t, \quad \forall t \in \mathcal{T}, \forall n \in \mathcal{N}.
\end{equation}
Each device has a RF-energy harvester to harvest energy from the remaining $(1-\rho^t_n)$ of signal $\mathbf{x^{\prime}}$. We consider a non-linear energy harvesting model \cite{7264986} and the harvested power at device $n$ is defined as
    \begin{equation} 
        \psi(P^t_n)=\frac{\frac{M}{1+\exp(-a P^t_{n}+b)}-M/(1+\exp(ab))}{1-1/(1+\exp(ab))}, 
    \label{equ:psi} 
    \end{equation} 
where $M$ is the maximum received power when the circuit is saturated; both $a$ and $b$ are constants specific to a given circuit. 
The harvested energy of device $n$ is 
\begin{equation}\label{equ:non-linear energy}
    H^t_n = \psi((1-\rho^t_n) P^t g^t_n)\tau.
\end{equation}
In an Uplink slot, each device splits its message into two parts, denoted as $s_{n,j}$, where $j \in \{1,2\}$. The received signal at the HAP is
\begin{equation}
    s_0 = \sum_{n=1}^{N}\sum_{j=1}^{2} \sqrt{p^t_{n,j}g^t_n}s_{n,j} + N_0,
\end{equation}
where $p^t_{n,j}$ is the transmit power of the $j$-th part of the message sent by device $n$. The HAP sets $p^t_{n,j}$ is to zero if slot $t$ is in Downlink mode. In addition, the total transmit power of device $n$ must be below its maximum transmit power $p_0$. The previous facts are formalized as follows:
\begin{equation}\label{uplink power}
    \sum_{j=1}^2 p^t_{n,j} \le p_0 \hat{I}^t, \quad \forall t \in \mathcal{T}, \forall n \in \mathcal{N}.
\end{equation}
The HAP decodes Uplink messages from users using SIC. To do so, it requires a decoding order which is defined as $[s_{n,j},\dots,s_{m,k}]$, where the first message to be decoded by the HAP is $s_{n,j}$. Define $\pi^k_{n,j}$ to denote the position of message $s_{n,j}$ in decoding order $k$, where message $s_{n,j}$ is decoded earlier than $s_{m,l}$ for order $k$ if $\pi^k_{n,j} > \pi^k_{m,j}$. 
Consider the decoding order $[s_{1,1},\dots,s_{N,1},s_{1,2},\dots,s_{N,2}]$.  Message $s_{1,1}$ is decoded earlier than $s_{2,1}$, where $\pi^k_{1,1} > \pi^k_{2,1}$.
The collection of all possible decoding orders is 
\begin{equation}\label{OriginalOrderSet}
\begin{aligned}
    \mathcal{K} = & \{[s_{1,1},\dots,s_{N,1},s_{1,2},\dots,s_{N,2}], \\
    & \dots,[s_{1,2},\dots,s_{N,2},s_{1,1},\dots,s_{N,1}],\},
\end{aligned}
\end{equation}
where each order is indexed by $k$. The number of decoding orders is $K=2N!$. 
Let $O^t_k$ be a binary variable that indicates whether decoding order $k$ is selected ($O^t_k=1$) by the HAP in time slot $t$. Only one order is selected if slot $t$ is in Uplink mode, i.e.,
\begin{equation}\label{order constraint}
    \sum_{k=1}^{2N!} O^t_k = \hat{I}^t, \quad \forall t \in \mathcal{T}.
\end{equation}
When order $k$ is selected, the HAP successfully decodes message $s_{n,j}$ if its SINR at the HAP, denoted as $\gamma^{t,k}_{n,j}$, is higher or equal to the threshold $\Gamma$. Formally,
\begin{equation}\label{Order SINR}
    \gamma^{t,k}_{n,j} = \frac{p^t_{n,j}g^t_n}{\sum_{\pi^k_{m,l}<\pi^k_{n,j}} p^t_{m,l}g^t_mU^t_{m,l}+N_0} \ge \Gamma.
\end{equation}
After that, the HAP removes message $s_{n,j}$ from the composite signal $x^{\prime}$. Let $U^t_{n,j}$ be a binary variable to indicate if the HAP decodes a message transmitted by device $n$ in slot $t$, where $U^t_{n,j} = 1$ means that the HAP successfully decoded a message sent by device $n$ in slot $t$. Formally, 
\begin{equation}\label{uplink SINR}
    \gamma^{t,k}_{n,1} + (2-O^t_k-U^t_{n,1})\Phi \ge \Gamma, \quad \forall t \in \mathcal{T},  \forall n \in \mathcal{N},  \forall k \in \mathcal{K},
\end{equation}
and
\begin{equation}\label{uplink SINR 2}
    \gamma^{t,k}_{n,2} + (2-O^t_k-U^t_{n,2})\Phi \ge \Gamma, \quad \forall t \in \mathcal{T},  \forall n \in \mathcal{N},  \forall k \in \mathcal{K}.
\end{equation}
In addition, we define 
\begin{equation}\label{uplink transmission}
    U^t_{n,j} \le \hat{I}^t, \quad \forall t \in \mathcal{T}, \forall n \in \mathcal{N}, \forall j \in {1,2},
\end{equation}
which ensures  devices transmit messages to the HAP only when slot $t$ is in Uplink mode. We also define 
\begin{equation}\label{order transmission}
    U^t_{n,j} \le \sum_{k=0}^{2N!}O^t_k, \quad \forall t \in \mathcal{T}, \forall n \in \mathcal{N}, \forall j \in {1,2},
\end{equation}
which ensures that devices transmit messages only when the HAP selects one order.  The number of uplink transmissions from device $n$ in slot $t$ is then calculated as $\hat{R}^t_n=U^t_{n,1}+U^t_{n,2}$.
%

%
In each slot, each device $n$ must have energy, meaning its energy level evolves over time as follows:
    \begin{equation}\label{energy constraint}
       E^{t+1}_n = E^t_n + H^t_n \breve{I}^t - \sum_{j=1}^{2} p^t_{n,j}\tau\ge 0.
    \end{equation}
We consider fairness between the amount of data transmitted in uplink and downlink. Define $\delta\in [0,1]$ as a weight that controls the number of uplink and downlink slots. 
The weighted sum-throughput over $T$ slots is calculated as
\begin{equation} \label{equ:weighted sum-throughput}
R =  \sum _{t=1}^T \sum _{n=1}^N ((1-\delta) \breve{R}^t_n + \delta \hat{R}^t_n).
\end{equation}
%
%
Note that $\breve{R}^t_n$ and $\hat{R}^t_n$ have the same magnitude. Specifically, observe that a device can either receive or transmit two messages in each slot. This means the maximum value of $\breve{R}^t_n$ and $\hat{R}^t_n$ is $2T$, where $T$ is the number of time slots.
Our aim is to maximize the weighted sum-throughput $R$ over $T$ slots. In Section~\ref{sec:MILP}, we first formulate a MILP, where it yields the optimal value of $R$.  A key limitation, however, is that it requires non-causal channel information.  To this end, in Section~\ref{sec:Q-Learning}, we propose a learning based approach when the HAP uses only causal channel information.

\section{Optimization Model}\label{sec:MILP}
To maximize $R$, we first formulate our problem as a Mixed Integer Non-Linear Program (MINLP). Define vector $\mathbf{v}=[\hat{I}^t,\breve{I}^t,D^t_n,\mu^t_c,\mu^t_n,\rho^t_n,U^t_n,p^t_{n,j},O^t_k]$ to contain all decision variables. The MINLP is formulated as
\begin{maxi!}|s|[3] 
{\mathbf{v}}{R}{\label{equ:MINLP}}{\text{(P1)} \quad} 
\addConstraint{}{\eqref{mode constraint},\eqref{downlink power},\eqref{downlink transmission}-\eqref{PS constraint},\eqref{uplink power},\eqref{order constraint},\eqref{uplink SINR}-\eqref{energy constraint}. \label{MINLP constraints}}{}{}
\end{maxi!}
In order to solve problem~\eqref{equ:MINLP} as an MILP, which can be solved more efficiently as compared to MINLP, the next subsection linearizes the non-linear terms $\psi(P^t_n)$, $\gamma_{c,n}^t$ and $\gamma_{n}^t$.
%

%
\subsection{Linearization}
First, we linearize Eq.~\eqref{equ:non-linear energy}. Define a set $\mathcal{S} = \{1,2,\dots,S\}$ of intervals. We use $P^L_s$ and $P^U_s$ to denote the lower and upper bound of interval $s$. If the received power $P^t_n$ falls within the interval $s$, we use a conversion efficiency $\eta_s$ to convert the received power, which is defined as
    \begin{equation}
        \eta_s = \frac{\psi(P^L_s)}{2 P^L_s}+\frac{\psi(P^U_s)}{2 P^U_s}.
        \label{eta_s}
\end{equation}
We use the binary variable $J^t_{s,n}$ to indicate that the received power of device $n$ falls in the $s$-th interval in slot $t$. The variable $J^t_{s,n}$ is set as follows:
    \begin{equation}\label{J}
        P^t_n \ge (J^t_{s,n}-1)\Phi + P^L_s, \quad \forall t \in \mathcal{T}, \forall n \in \mathcal{N}, \forall s \in \mathcal{S}, 
    \end{equation}
    \begin{equation}\label{J 2}
         P^t_n \le P^U_s+(1-J^t_{s,n})\Phi, \quad \forall t \in \mathcal{T}, \forall n \in \mathcal{N}, \forall s \in \mathcal{S}. 
    \end{equation}
When the received power of device $n$ falls in the $s$-th interval in slot $t$, where $P^L_s\le P^t_n \le P^U_s$, the variable $J^t_{s,n}$ is set to one. Otherwise, $J^t_{s,n}$ is set to zero so that the inequality~\eqref{J} holds. In addition, each $J^t_{n,s}$ is set to zero if slot $t$ is in Uplink mode, where 
    \begin{equation}
        \sum_{s=1}^S J^t_{s,n} \le \breve{I}^t, \quad \forall t \in \mathcal{T}, \forall n \in \mathcal{N}.
    \end{equation}
Define $M^t_{s,n}$ as the harvested energy of device $n$ in slot $t$ if the received power $P^t_n$ falls within the interval $s$. We determine the converted power $M^t_{s,k}$ as follows:
    \begin{equation}
       M^t_{s,n} \le \eta_s P^t_n, \quad \forall t \in \mathcal{T}, \forall n \in \mathcal{N}, \forall s \in \mathcal{S},
       \label{M^t_{s,k}}
    \end{equation}
    \begin{equation}
       M^t_{s,n} \ge \eta_s P^t_n-(1-J^t_{s,n})\Phi, \quad \forall t \in \mathcal{T}, \forall n \in \mathcal{N}, \forall s \in \mathcal{S},
       \label{M^t_{s,k} 1}
    \end{equation}
    \begin{equation} 
    0 \le M^t_{s,n} \le J^t_{s,n}\Phi, \quad \forall t \in \mathcal{T}, \forall n \in \mathcal{N}, \forall s \in \mathcal{S}.
    \label{M^t_{s,k} 2}
    \end{equation}
Specifically, the output power $M^t_{s,n}$ is set to $\eta_s P^t_n$ when $J^t_{s,n}$ is set to one. Similarly, if $J^t_{s,n}$ is zero, the output power $M^t_{s,n}$ is set to zero. 
The harvested energy of device $n$ in slot $t$ is then calculated as $H^t_n = \sum_{s=1}^{S} M^t_{s,n}\tau$. 
%

In downlink slots, the SINR of common message $W^c$ and private message $W^p_n$, see Eq.~\eqref{common SINR 2} and \eqref{private SINR 2}, involve a product of two real decision variables. To linearize this, we apply McCormick envelopes\footnote{Note that the resulting MILP yields the lower bound of the optimal solution.  This bound can be improved by dividing up the domain of decision variables and applying linearization in each sub-divided domain.}. 
Let $\mu^U_c, \mu^L_c,\mu^U_n, \mu^L_n, \rho^U, \rho^L$ denote the upper and lower bound value of $\mu^t_c, \mu^t_n$ and $\rho^t_n$, respectively. Define a decision variable $\omega^t_c$ to replace the product of $\mu^t_c$ and $\rho^t_n$. Inequality~\eqref{equ:gamma_cn} is then revised to
    \begin{equation}
        \gamma_{c,n}^t = \frac{\omega^t_{c,n} P g^t_n}{\sum_{n=1}^N \mu_n \rho^t_n P g^t_n + N_o} \ge \Gamma.
    \end{equation}
Specifically, the term $\omega^t_c$ is constrained by 
    \begin{equation} \label{omega 1}
        \omega^t_{c,n} \ge \mu^L_c\rho^t_n + \mu^t_c\rho^L - \mu^L_c\rho^L, \quad \forall t \in \mathcal{T}, \forall n \in \mathcal{N},
    \end{equation}
    \begin{equation}
        \omega^t_{c,n} \ge \mu^U_c\rho^t_n + \mu^t_c\rho^U - \mu^U_c\rho^U, \quad \forall t \in \mathcal{T}, \forall n \in \mathcal{N},
    \end{equation}
    \begin{equation}
        \omega^t_{c,n} \le \mu^U_c\rho^t_n + \mu^t_c\rho^L - \mu^U_c\rho^L, \quad \forall t \in \mathcal{T}, \forall n \in \mathcal{N},
    \end{equation}
    \begin{equation} \label{omega 2}
        \omega^t_{c,n} \le \mu^L_c\rho^t_n + \mu^t_c\rho^U - \mu^L_c\rho^U, \quad \forall t \in \mathcal{T}, \forall n \in \mathcal{N}.
\end{equation}
The above constraints create a convex hull that encapsulates the value of $\mu^t_c\cdot \rho^t_n$.
As for the product of $\mu^t_c$ and $\rho^t_n$, define a decision variable $\omega^t_n$. Inequality~\eqref{equ:gamma_n} is then revised to
    \begin{equation}
        \gamma_{n}^t = \frac{\omega^t_n P g^t_n}{\sum _{m\neq n} \mu^t_m \rho^t_n P g^t_m + N_o} \ge \Gamma.
    \end{equation}
In particular, the term $\omega^t_n$ is constrained by 
    \begin{equation}\label{omega 3}
        \omega^t_n \ge \mu^L_n\rho^t_n + \mu^t_n\rho^L - \mu^L_n\rho^L, \quad \forall t \in \mathcal{T}, \forall n \in \mathcal{N},
    \end{equation}
    \begin{equation}
        \omega^t_n \ge \mu^U_n\rho^t_n + \mu^t_n\rho^U - \mu^U_n\rho^U, \quad \forall t \in \mathcal{T}, \forall n \in \mathcal{N},
    \end{equation}
    \begin{equation}
        \omega^t_n \le \mu^U_n\rho^t_n + \mu^t_n\rho^L - \mu^U_n\rho^L, \quad \forall t \in \mathcal{T}, \forall n \in \mathcal{N},
    \end{equation}
    \begin{equation}\label{omega 4}
        \omega^t_n \le \mu^L_n\rho^t_n + \mu^t_n\rho^U - \mu^L_n\rho^U, \quad \forall t \in \mathcal{T}, \forall n \in \mathcal{N}.
    \end{equation}
In our model, the power coefficient $\mu^t_c$ and $\mu^t_n$ are non-negative and constrained by the maximum transmit power of the HAP. Hence, we set the value of $\mu^t_c$ and $\mu^t_n$ to be in the range $[0,1]$ (in Watt), respectively. Therefore, we have 
    \begin{equation}\label{upper bound}
        0 \le \mu^t_c \le 1, \quad \forall t \in \mathcal{T}, \forall n \in \mathcal{N},
    \end{equation}
    \begin{equation}
        0 \le \mu^t_n \le 1, \quad \forall t \in \mathcal{T}, \forall n \in \mathcal{N}.
    \end{equation}
In addition, for the power splitting ratio $\rho^t_n$, we have
    \begin{equation}\label{lower bound}
        0 \le \rho^t_n \le 1, \quad \forall t \in \mathcal{T}, \forall n \in \mathcal{N}.
    \end{equation}
Rewriting inequality~\eqref{omega 1}-\eqref{omega 2} and \eqref{omega 3}-\eqref{omega 4}, we have 
    \begin{equation}\label{equ:omega 1}
       \rho^t_n \ge \omega^t_{c,n} \ge 0, \quad \forall t \in \mathcal{T}, \forall n \in \mathcal{N},
    \end{equation}
    \begin{equation}
       \mu^t_c \ge \omega^t_{c,n} \ge \rho^t_n + \mu^t_c - 1, \quad \forall t \in \mathcal{T}, \forall n \in \mathcal{N},
    \end{equation}
    \begin{equation}
       \rho^t_n \ge \omega^t_n \ge 0, \quad \forall t \in \mathcal{T}, \forall n \in \mathcal{N},
    \end{equation}
    \begin{equation}\label{equ:omega 2}
       \mu^t_n \ge \omega^t_n \ge \rho^t_n + \mu^t_n - 1, \quad \forall t \in \mathcal{T}, \forall n \in \mathcal{N}.
    \end{equation}

Define the vector $\mathbf{v^{\prime}}=[J^t_{s,n},M^t_{s,n},\omega^t_{c,n},\omega^t_n]$ to contain  all decision variables used in the aforementioned linearization. We thus have the following MILP:
\begin{maxi!}|s|[3] 
{\mathbf{v},\mathbf{v^{\prime}}}{R}{\label{equ:MILP}}{\text{(P2)} \quad} 
\addConstraint{}{\eqref{MINLP constraints},}{\eqref{J}-\eqref{M^t_{s,k} 2},}{\eqref{upper bound}-\eqref{equ:omega 2}. \label{MILP constraint}}
\end{maxi!}
The computational complexity of MILP~\eqref{equ:MILP} is related to its number of variables and constraints.  Specifically, we have the following result:
\begin{prop} \label{solution space}
Assume there are $N$ devices and $T$ slots. Then the solution space has size $\sum_{i=0}^T \binom{T}{i} \left(\sum_{j=1}^{N} \binom{2N}{j} (2N)!\right)^i$.
\end{prop}
\begin{proof}
For $T$ slots, the HAP has $\binom{T}{i}$ choices of Uplink slots, where $i$ denotes the number of Uplink slots over $T$ slots. In each Uplink slot, there are $(2N)!$ choices of decoding orders. When an order is selected, devices have zero or at most $2N$ messages transmitted. If the HAP selects $j$ devices to transmit in each data slot, there are $\binom{2N}{j}$ choices of transmitting devices.
These facts imply the size of solution space is $\sum_{i=0}^T \binom{T}{i} \left(\sum_{j=1}^{N} \binom{2N}{j} (2N)!\right)^i$.
\end{proof}
Proposition~\eqref{solution space} implies that MILP~\eqref{equ:MILP} becomes computationally intractable for large-scale networks. 
\subsection{Reduced Order Set}
We observe that for a decoding order, if we swap the position of two messages from same device, constraint~\eqref{uplink  SINR} and \eqref{uplink  SINR 2} still hold. 
To illustrate, assume the HAP uses $[s_{1,1},s_{1,2},\dots,s_{N,1},s_{N,2}]$ to decode Uplink messages and the transmit power of message $s_{1,1}$ is $p^{\prime}$. The SINR of message $s_{1,1}$ is 
\begin{equation}
    \gamma^{t,k}_{n,j} = \frac{p^{\prime}g^t_n}{\sum_{\pi^k_{m,l}<\pi^k_{1,1}} p^t_{m,l}g^t_mU^t_{m,l}+N_0}.
\end{equation}
According to constraint~\eqref{uplink power}, the transmit power of messages is a decision variable, so the transmit power assigned to message $s_{1,1}$ can also be assigned to $s_{1,2}$. Thus, message $s_{1,2}$ has the same SINR if the HAP uses decoding order $s_{1,2},[s_{1,1},\dots,s_{N,1},s_{N,2}]$.
Given the observation, we reduce the value of $K$ by removing decoding orders that only differ in the position of messages from the same device. The size of the order set then becomes $K = \frac{(2N)!}{2^N}$. Eq.~\eqref{order constraint} is revised to
\begin{equation}\label{order constraint 2}
    \sum_{k=1}^{\frac{(2N)!}{2^N}} O^t_k = \hat{I}^t.
\end{equation}
We thus have the following MILP with reduced order set, namely {\em MILP-Re}:
\begin{maxi!}|s|[3] 
{\mathbf{v},\mathbf{v^{\prime}}}{R}{\label{equ:MILP Reduce}}{\text{(P3)} \quad}
\addConstraint{}{\eqref{mode constraint},\eqref{downlink power},\eqref{downlink transmission}-\eqref{PS constraint},\eqref{uplink power},\eqref{uplink SINR}-\eqref{energy constraint},\eqref{order constraint 2}}{}
\addConstraint{}{\eqref{J}-\eqref{M^t_{s,k} 2},\eqref{upper bound}-\eqref{equ:omega 2}.}{}
\end{maxi!}

Solving problem (P2) or (P3) yields the optimal weighted sum-throughput over a planning horizon. A key limitation of this problem is that it requires non-causal channel state information. To this end, in the following section, we present a learning based approach to maximize the weighted sum-throughput over $T$ slots using causal channel information. Note that MILP~\eqref{equ:MILP} and \eqref{equ:MILP Reduce} can be used as a benchmark against solutions that do not have channel state information. 
%

\section{A reinforcement learning-based approach}\label{sec:Q-Learning}
We first formulate an Markov Decision Process (MDP) \cite{8714026}, where an MDP is used to model a sequential decision making process. After that, we propose a Q-learning based approach to maximize the weighted sum-throughput using causal channel state information.  
Note the application of Q-learning is not straightforward as in addition to the mode of time slots, we also have to consider the subset of devices and their order of uplink/downlink transmissions in each time slot.
Apart from that, we emphasize that other reinforcement learning solutions can also be used.  However, given that the application of reinforcement learning to our problem is an open problem, it is important to {\em first} establish how it can be applied to our problem before applying more computationally expensive reinforcement learning solutions, which belongs in a future work\footnote{Note, our results show that Q-learning managed to achieve 90\% of the optimal result.  Hence, it is an open question whether the additional computational complexity of methods such as deep reinforcement learning will yield any significant gains.}.

\subsection{MDP}
Define a MDP as a tuple $\{ \mathcal{S}, \mathcal{A}, \mathcal{T}, \mathcal{R} \}$, which records the set of states $\mathcal{S}$, set of actions, $\mathcal{A}$, set of rewards $\mathcal{R}$, and transition probability from one state to another $\mathcal{T}$.  Formally, 
\begin{enumerate}
\item \textbf{State} $\mathcal{S}$. Define $\mathbf{s}^t \in \mathcal{S} $ as the state of the system in slot $t$, where $\mathbf{s}^t =[s^t_1,\dots,s^t_n]$ contains the state of all devices. Each $s^t_n$ corresponds to the maximum received power from device $n$ to the HAP, which is the product of the residual energy at device $n$ and the channel gain between device $n$ and the HAP. Let $\mathcal{L} = \{0,\epsilon_0, \dots, L\epsilon_0\}$ denote the set of all possible power levels that devices can have, where $L$ represents the number of power levels. As a result, in our MDP, we have a finite state space $\mathcal{S} = \{ [0, \dots,0] , [0, \dots, \epsilon_0], \dots, [\epsilon_0, \dots, \epsilon_0], \dots, [L\epsilon_0, \dots, L\epsilon_0] \}$, where $\epsilon_0 = 1 \ \mu W$. The size of the state space is defined as $N_s$, where  $N_s = L^N$.
\item \textbf{Action} $\mathcal{A}$. We define an action space $\mathcal{A} = \{0,1\}$, where $a^t \in \mathcal{A}$ is the action in slot $t$. In particular, we have $a^t = 0$ if the HAP selects Downlink mode in slot $t$; conversely, we have $a^t = 1$ for Uplink mode.
\item \textbf{Transition probability} $\mathcal{T}$. We consider a model-free approach, meaning the transition probability is unknown. 
\item \textbf{Reward} $\mathcal{R}$. The reward $r(a^t)$ is defined as the weighted sum-throughput when taking action $a^t$ in slot $t$, which is calculated according to Eq.~\eqref{equ:weighted sum-throughput}.
\end{enumerate}
Our aim is to determine a policy that maximizes the weighted sum-throughput $R$, see Eq.~\eqref{equ:weighted sum-throughput}.  The said policy selects an action $a^t$ for a given state $\mathbf{s}^t$ that maximizes the expected weighted sum-throughput. Define a policy as $\pi(\mathbf{s}^t)$, where $\pi: \mathcal{S} \rightarrow \mathcal{A}$. Formally, the problem is
\begin{equation}
    \max_{\pi}\lim_{T \to \infty} \frac{1}{T} \mathbb{E} \left[ \sum_{t=1}^{\infty} R \right].
\end{equation}
\subsection{Q-Learning}
The HAP/agent uses Q-learning to find the optimal policy by maintaining a Q-table, which contains so called Q-values indicating the expected reward of each state-action pair. Given the Q-table, the HAP selects an action according to the environment of the system in each time slot.  Then in each time slot, it determines the optimal system parameters, such as power splitting ratio, power coefficient and uplink decoding order using linear programs, which will be presented in Section~\ref{sec:DLP} and \ref{sec:ULP}. 
Note that in practice, the HAP/agent only requires the channel gain to/from each device and the energy level of devices.  Both information can be collected at the start of each time slot.

Let $Q(\mathbf{s}^t,a^t)$ be the said Q-table, which contains Q-values that indicate the expected discounted reward for the state-action pair $(\mathbf{s}^t,a^t)$. Specifically, the HAP takes an action $a^t$ in state $\mathbf{s}^t$ and observes $Q(\mathbf{s}^t,a^t)$, where it then updates its Q-table according to Bellman's equation \cite{8714026}:
\begin{equation}\label{equ:update Q}
\begin{aligned}
Q(\mathbf{s}^{t},a^{t}) & = (1-\alpha) Q(\mathbf{s}^{t-1},a^{t-1}) \\ 
& + \alpha(r(\mathbf{s}^t,a^t)+ \gamma \max Q(\mathbf{s}^{t+1},a^{t+1})),
\end{aligned}
\end{equation} 
where $\gamma \in [0,1]$ is a discount factor, $\alpha \in [0,1]$ is a learning ratio, and $r(\mathbf{s}^t,a^t)$ represents the immediate reward for the state-action pair $(\mathbf{s}^t,a^t)$.

We now present the main body of the Q-Learning approach, see Algorithm~\ref{algo:Q}. The HAP first initializes its Q-table arbitrarily. Define $k \in \mathcal{K}$ as the index of a training episode, where $\mathcal{K} = \{1,2,\dots,K\}$. In each training episode, the HAP selects an action using {\em $\epsilon$-greedy} \cite{8714026}. Specifically, in line \ref{e-greedy start}-\ref{e-greedy end}, the HAP executes a random action with a probability $\epsilon$, where it selects action $a^t$ with the highest $Q(\mathbf{s}^t,a^t)$ with probability $1-\epsilon$. In addition, the value of $\epsilon$ decays from one to zero over time, see line \ref{decay}. If the HAP selects $a^t = 0$, see line \ref{downlink start}-\ref{downlink end}, it then determines the power splitting ratio ${\rho}^t_n$ and the power coefficient $\mu^t_c$ and $\mu^t_n$ using {\em DLP} as outlined in Section~\ref{sec:DLP}. When the HAP selects $a^t = 1$, see line \ref{uplink start}-\ref{uplink end}, it determines the decoding order and the transmit power $p^t_{n,j}$ of each transmission as per {\em ULP}, see Section~\ref{sec:ULP}. After that, as shown in line \ref{update}, the HAP updates its Q-table using reward $R^t_n$ according to Eq.~\eqref{equ:update Q}.
\begin{algorithm}[htbp]
\caption{{Q-Learning}.}  
\hspace*{\algorithmicindent} \textbf{Input:} $\mathcal{T}, \mathcal{K},\epsilon$\\
\hspace*{\algorithmicindent} \textbf{Output:} $\mathcal{Q}$
\label{algo:Q}   
\begin{algorithmic}[1]  
\State $\mathcal{A} = \{0,1\}, \mathcal{Q} = \{\}$
\For {each episode $k \in \mathcal{K}$}
\For {each slot $t \in \mathcal{T}$} 
\State $\mathbf{s}^t = ${\em CollectState()}
\State $\hat{z}$ = {\em GenerateRandom}($[0,1]$)
\If {$\hat{z} \le \epsilon$ } \label{e-greedy start}
\State Randomly select an action $a^t \in \mathcal{A}$.
\Else
\State $a^t = \argmax\limits_{a^t\in\mathcal{A}} \mathcal{Q}(\mathbf{s}^t)$
\EndIf\label{e-greedy end}
\State {\em Decay($\epsilon$)}\label{decay}
\If {$a^t = 0$ } \label{downlink start}
\State $R^t_n$ = {\em DLP}$(\mathbf{s}^t)$
\EndIf\label{downlink end}
\If {$a^t = 1$ }\label{uplink start}
\State $R^t_n$ = {\em ULP}$(\mathbf{s}^t)$
\EndIf\label{uplink end}
\EndFor
\State {\em UpdateQtable}$(R^t_n,\mathcal{Q}(\mathbf{s}^{t},a^{t}))$ \label{update}
\EndFor
\State \Return $\mathcal{Q}$
\end{algorithmic}  
\end{algorithm}	
\subsection{{\em Downlink Linear Program (DLP)}} \label{sec:DLP}
In each Downlink slot, the HAP solves an LP to determine the power splitting ratio ${\rho}^t_n$, the power coefficient $\mu^t_c$ and $\mu^t_n$ that maximizes the number of {\em Dowlink} messages in the current slot. 
The LP is formulated as
\begin{maxi!}|s|[3]  
{[{\rho}^t_n,\mu^t_c,\mu^t_n]}{\sum_{n=1}^{N} C^t_n +D^t_n}{\label{equ:DLP}}{} \addConstraint{}{\eqref{common SINR},\eqref{private SINR},\eqref{downlink transmission},\eqref{downlink transmission 1},\eqref{PS constraint},\eqref{equ:omega 1}-\eqref{equ:omega 2}. \label{DLP}}{}{}
\end{maxi!} 
In constraint~\eqref{downlink transmission} and \eqref{downlink transmission 1}, we have $\breve{I}^t=1$, which set the current slot to Downlink mode. 
\subsection{{\em Uplink Linear Program (ULP)}} \label{sec:ULP}
In each Uplink slot, the HAP solves an LP to decide the decoding order and the transmit power $p^t_{n,j}$ of each transmission that yields the maximum number of uplink transmissions. The LP is formulated as
\begin{maxi!}|s|[3]  
{[p^t_{n,j},O^t_k]}{\sum_{n=1}^{N}\sum_{j=1}^{2} U^t_{n,j}}{\label{equ:ULP}}{} 
\addConstraint{}{\eqref{uplink power},\eqref{order constraint},\eqref{uplink SINR}-\eqref{order transmission},\eqref{upper bound}-\eqref{lower bound}. {}{}}
\end{maxi!} 
In constraint~\eqref{uplink power}, \eqref{order constraint}, and \eqref{uplink transmission}, we set $\hat{I}^t=1$ to indicate slot $t$ is in Uplink mode.

%
%
\section{Evaluation}\label{sec:Eval}
Our evaluation uses a simulator written in Python and a computer with an Intel i7-7700 CPU @3.6GHz with 16 GB of memory. We solve MILP (P2) and (P3) using Gurobi.  The HAP has a transmit power of $2$ W \cite{powercast}.  Devices are placed $5$ m away from the HAP; at this distance, devices are able to harvest energy from the HAP using the harvester from Powercast~\cite{powercast}. The SINR threshold is $2$ dB.  The RF-energy harvester at devices,  as per \cite{7264986}, has a maximum received power of $0.024$ W, with parameter value $\alpha_j=150$ and $\beta_j=0.014$. The battery size of each device is set to $2500$ mAh \cite{5522465}
The packet size is fixed to $250$ Kb \cite{4460126}. The circuit noise power is $10^{-13}$ W/Hz.  The channel bandwidth is $20$ MHz, which is used by WiFi. 
We note that if the weight of downlink transmission is higher than $0.5$, the HAP will always select Downlink mode. As a result, the weight $\alpha$ of Downlink sum-throughput in Eq.~\eqref{equ:weighted sum-throughput} is set to $0.4$.  
Further, we do not consider wireless errors.  Hence, our results are upper bound in terms of the number of messages transmitted/received by devices/HAP.
Table~\ref{tbl:parameter} lists our parameter values. 
\begin{table}[htbp] 
\begin{center} 
\caption{A summary of parameter values.} 
\label{tbl:parameter} 
\begin{tabular}{l l} 
\hline 
\bf Parameter & \bf Value \\ \hline 
Maximum HAP transmit power ($P_{max}$) & $2$\\ 
Distance of device $n$ to the HAP ($d_{n}$) & $5$ m \\ 
Energy to transmit a packet ($\epsilon$) & $4.75$ $\mu J$ \\ 
SINR threshold $\Gamma$ & $2$ dB \\ 
Maximum received power ($M_j$) & $0.024$ W \\ 
Parameter of non-linear model ($a_j)$ & $150$ \\ 
Parameter of non-linear model ($b_j$) & $0.014$ \\ 
Noise power & $10^{-13}$ W/Hz \\ 
Channel bandwidth & $20$ MHz \\
Number of runs & $1000$ \\ \hline 
\end{tabular} 
\end{center} 
\end{table}

We benchmark against three other solutions:
\begin{itemize}  
\item {\em Optimal solution or MILP}. The HAP uses non-causal channel gain information to solve MILP \eqref{equ:MILP}. Hence, it obtains the optimal mode for each slot, as well as power splitting ratios, power coefficients in Downlink slots, and device transmit power in Uplink slots.  
\item {\em MILP with Reduced Order Set (MILP-Re)}. The HAP solves MILP \eqref{equ:MILP Reduce} to obtain the aforementioned quantities.  
\item{\em Random (Rand)}. The HAP randomly selects a mode for each slot.  Then in Downlink slots, the HAP uses {\em DLP}. As for Uplink slots, it uses {\em ULP}.  
\item{\em Time Division Duplex (TDD)}.  The HAP alternates between a Downlink and Uplink slot and applies {\em DLP} and {\em ULP} accordingly.  
\end{itemize}

We conduct $1000$ runs for each experiment, where each run contains $10$ slots. In each experiment, we record the average weighted sum-throughput as defined in Eq.~\eqref{equ:weighted sum-throughput}.
%

%
\subsection{Decay factor}
\begin{figure}[htbp] 
\centering 
\includegraphics[width=0.8\linewidth]{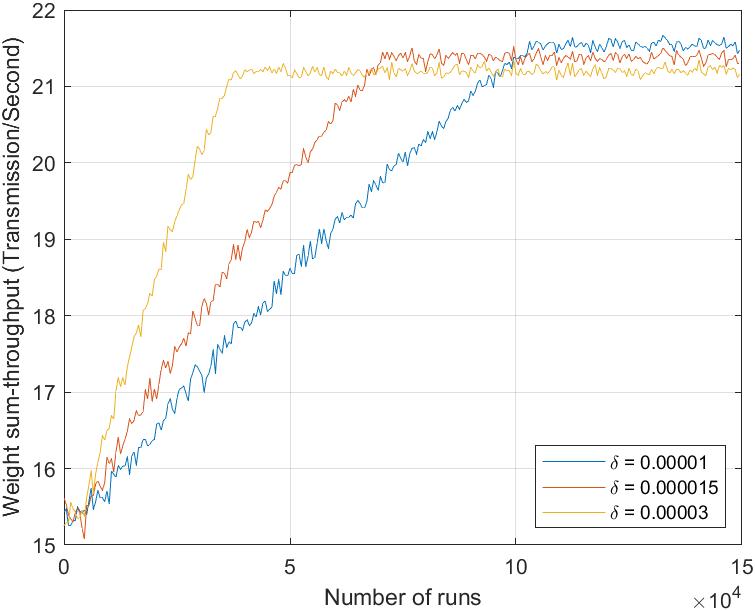} 
\caption{Impact of different decay factors.} 
\label{fig:4.0.1} 
\end{figure}

First, we study the decay factor $\delta$ of {\em Q-Learning}. 
Referring to Fig.~\ref{fig:4.0.1}, for higher $\delta$ values, {\em Q-Learning} achieves a higher weighted sum-throughput. The weighted sum-throughput of {\em Q-Learning} with $\delta = 0.00001$ after convergence is $1.6\%$ higher than $\delta = 0.00003$. In addition, the convergence time of {\em Q-Learning} increases when $\delta$ increases. Specifically, the {\em Q-Learning} agent with $\delta=0.00003$ converges after $40000$ runs, where the agent with $\delta=0.00001$ requires approximately $105000$ runs before it converges. This is because $\epsilon$ decreases to zero faster when a higher $\delta$ is used, where the agent at the HAP has less chance to explore actions before converging to a mode. On the other hand, when a smaller $\delta$ is used, the value of $\epsilon$ decays slower, where the {\em Q-Learning} agent will have more chance to explore different modes. 
 
\subsection{Distance} \label{} 
\begin{figure}[htbp]  
\centering  
\includegraphics[width=0.8\linewidth]{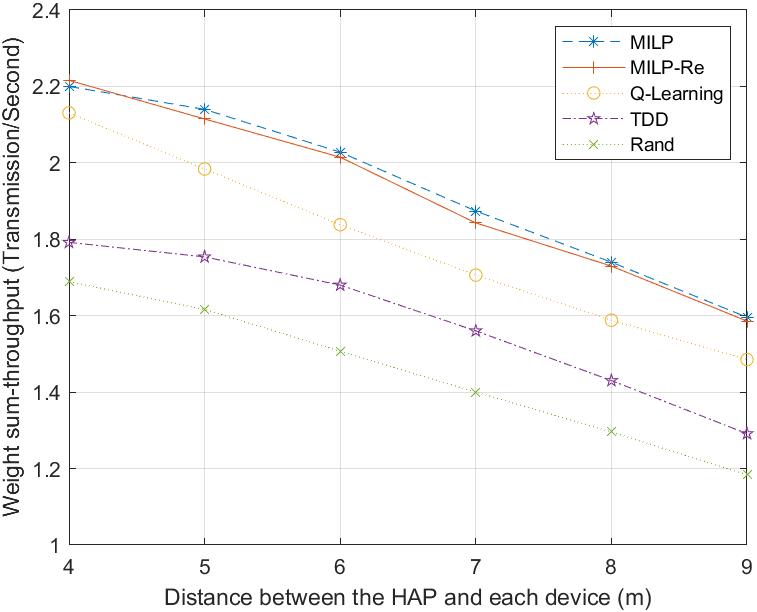}  
\caption{ Weighted sum-throughput for varying HAP and device distance.}  
\label{fig:4.1.1}  
\end{figure}

We first vary the distance between the HAP and each device from $4$ m to $9$ m. As shown in Fig.~\ref{fig:4.1.1}, the weighted sum-throughput of all approaches decreases.
This is because the channel gain becomes smaller when devices are placed further away from the HAP according to Eq.~\eqref{channel}. As a result, devices harvest less energy in each Downlink slot, which results in more Downlink slots required by devices in order to harvest sufficient energy for Uplink slots.  
The same reason applies to {\em Q-Learning}, where the energy harvested by devices decreases when the distance between the HAP and each device increases, which results in fewer uplink slots. 
%

%
%

%
As per Fig.~\ref{fig:4.1.1}, the weighted sum-throughput of {\em MILP} is around $1.4\%$ on average higher than {\em MILP-Re}. This is because when the HAP uses {\em MILP-Re}, it selects the decoding order from a reduced order set.  {\em MILP-Re} may select an order that results in fewer uplink transmissions or has a higher energy consumption in Uplink slots. 
%
From Fig.~\ref{fig:4.1.1}, the weighted sum-throughput of {\em MILP} is higher than {\em Q-Learning}, where the maximum gap occurs when devices are placed $6$ m away from the HAP.  This is due to the fact that {\em MILP} has non-causal channel information for selecting a mode, whereas {\em Q-Learning} has channel information of current and past slots.  Moreover, {\em Q-Learning} may select Uplink mode in slots that should be in Downlink mode to ensure more messages in future Uplink slots. As a result, {\em Q-Learning} achieves fewer uplink transmissions than {\em MILP} when the distance between the HAP and each device is $6$ m.  
%
The weighted sum-throughput of {\em Q-Learning} is $11.9\%$ higher than {\em TDD} when the distance between the HAP and each device increases from $4$ to $9$ m. This is reasonable because {\em Q-Learning} learns when to select Downlink mode using the channel condition and the residual energy at devices. However, the mode is pre-determined for {\em TDD}, where devices may accumulate too much energy with no chance for uplink transmission, or lack of energy for uplink transmissions.
%
 
%





\subsection{Maximum HAP transmit power} 
\begin{figure}[htbp]  
\centering  
\includegraphics[width=0.8\linewidth]{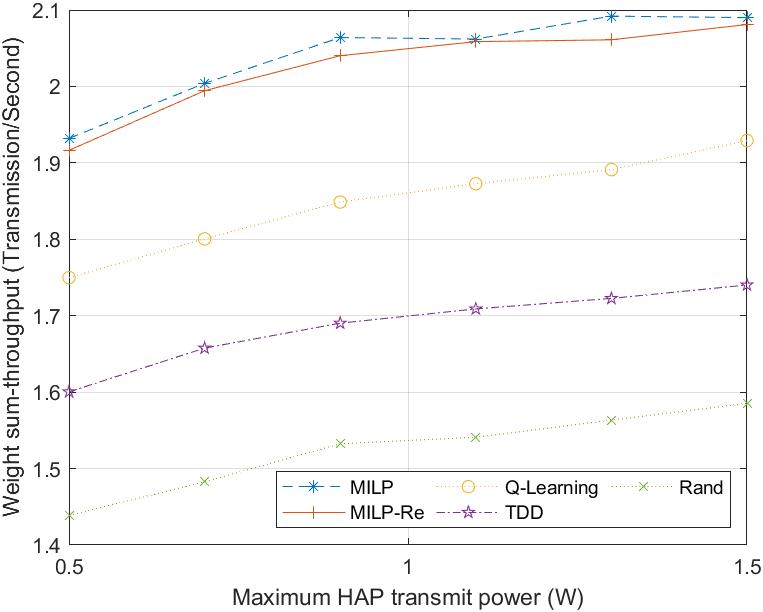}  
\caption{Impact of HAP transmit power on weighted sum-throughput.}  
\label{fig:4.2.1}  
\end{figure}

We vary the maximum transmit power of the HAP from $0.5$ to $1.5$ W. As shown in Fig~\ref{fig:4.2.1}, the weighted sum-throughput of all approaches increases. 
The weighted sum-throughput of {\em MILP} and {\em Q-Learning} increases by around $8.8\%$ and $9.4\%$ when the maximum transmit power of the HAP increases from $0.5$ to $1.5$ W.
This is because devices harvest more energy in each Downlink slot if the HAP uses a higher transmit power. Thus, devices are more likely to have sufficient energy in uplink slots, which results in more uplink transmissions. 

%
As shown in Fig.~\ref{fig:4.2.1}, the maximum gap between {\em MILP} and {\em Q-Learning} occurs when the HAP uses a maximum transmit power of $0.9$ W; the weighted sum-throughput of {\em MILP} is around $9\%$ higher than {\em Q-Learning}. This is because {\em MILP} utilizes non-causal channel information to decide the mode of slots. By contrast, {\em Q-Learning} only uses channel information in current and past slots. As a result, the number of uplink transmissions attained by {\em MILP} is higher than {\em Q-Learning}.
%
%
As per Fig.~\ref{fig:4.2.1}, the weighted sum-throughput of {\em Q-Learning} is on average $10.7\%$ and $21.2\%$ higher than {\em TDD} and {\em Rand}, respectively. This is because {\em Q-Learning} determines the mode according to the maximum received power of uplink transmissions of devices. When the residual energy at devices is high, the HAP will select more Uplink slots. However, the mode is pre-determined for {\em TDD} and random for {\em Rand}, where devices may have insufficient chance or lack of energy for uplink transmissions. Thus, the HAP with {\em Q-Learning} receives more uplink transmissions than {\em TDD} and {\em Rand}
%

%

\subsection{SIC threshold} 
\begin{figure}[htbp]  
\centering  
\includegraphics[width=0.8\linewidth]{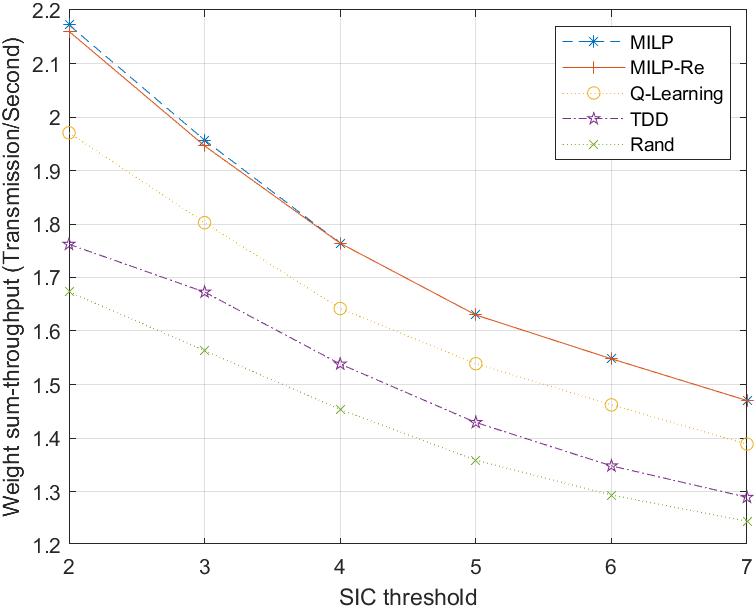}  
\caption{Weighted sum-throughput versus SIC threshold $\Gamma$.}  
\label{fig:4.4.1}  
\end{figure}

We vary the SIC decoding threshold from $2$ to $7$ (in dB). As shown in Fig.~\ref{fig:4.4.1}, the weighted sum-throughput of {\em MILP} decreased by $47.7\%$ when the SIC decoding threshold increases from $2$ to $7$. This is due to the higher transmit power required for successful SIC decoding, which results in devices consuming more energy. 
%
The same reason applies to {\em Q-Learning}. As shown in Fig.~\ref{fig:4.4.1}, the weighted sum-throughput of {\em Q-Learning} decreases by $41.8\%$, respectively, when the SIC decoding threshold increases from $2$ to $7$. 
%
 
%
As shown in Fig.~\ref{fig:4.4.1}, the weighted sum-throughput of {\em MILP} is $7.2\%$ higher than {\em Q-Learning} when the SIC decoding threshold is two, and remains approximately the same when the threshold increases to seven. This is reasonable because {\em MILP} uses non-causal channel information to determine a mode, whereas {\em Q-Learning} utilizes channel information of current and past slots. When using {\em Q-Learning}, the HAP may select Uplink mode instead of Downlink to ensure more uplink transmissions in future time slots. 
%
On the other hand, the weighted sum-throughput of {\em Q-Learning} is around $13.9$ on average higher than {\em Rand} when the SIC decoding threshold increases from two to seven. This is because {\em Q-Learning} learns to select the optimal mode for a given system state. 
However, the mode in each slot is random for {\em Rand}, which may cause devices to accumulate excessive amount of energy, or lack of chance for uplink transmission.  For these reasons, {\em Q-Learning} has better performance than {\em TDD} and {\em Rand}.
%
 
%
%

 
%
%



%
\subsection{Device Numbers} 
\begin{figure}[htbp]  
\centering  
\includegraphics[width=0.8\linewidth]{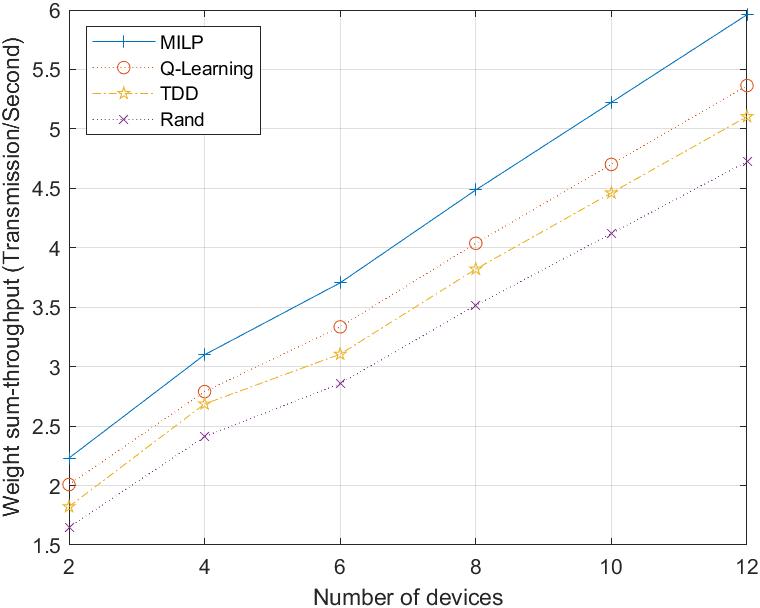}  
\caption{ Weighted sum-throughput versus the number of devices.}  
\label{fig:4.5.1}  
\end{figure}

We vary the number of devices from two to twelve.  Recall that for ULP, see Eq.~\eqref{OriginalOrderSet}, there are $(2N)!$ decoding orders in each uplink slot.  Hence, ULP becomes computationally intractable for large number of devices.  Thus, in this section, we reduce the computational complexity by fixing the decoding order as per the channel gain of devices. Specifically, messages sent by a device with a higher channel gain will be decoded earlier.   
From Fig.~\ref{fig:4.5.1}, the weighted sum-throughput of all approaches increases. This is because more devices are able to transmit in each uplink slot, which results in more uplink transmissions. 
%
%

%
According to Fig.~\ref{fig:4.5.1}, the weighted sum-throughput of {\em MILP} is higher than {\em Q-Learning}.
The reason is that {\em MILP} has non-causal channel information. It is possible for {\em Q-Learning} to select Uplink mode for a slot when in fact a Downlink mode is more suitable in order to save energy for future slots so that more devices can transmit messages in Uplink slots. As a result,  {\em MILP} selects more downlink slots than {\em Q-Learning}.
%
The weighted sum-throughput of {\em Q-Learning} is on average $6.4\%$ and $16.6\%$ higher than {\em TDD} and {\em Rand}, respectively, when the number of devices increases from two to twelve. This is because {\em Q-Learning} learns when to select {\em Downlink} mode using channel condition and the residual energy of devices. However, the mode is pre-determined for {\em TDD} and random for {\em Rand}, where devices may accumulate too much energy with no chance for uplink transmissions, or they lack energy for uplink transmissions. 
%

%
%
 



\section{Conclusion}\label{sec:conclusion}
This paper presents the first study on a RSMA-based IoT network that uses a mode based time structure.  It addresses a novel problem that requires a solution to determine whether a time slot is used for downlink or uplink transmissions. 
It outlines the first MILP that can be used to decide the optimal solution, namely the mode in each slot, the transmit power of each packet, the power splitting ratio of each device, and the decoding order in uplink slots. 
Further, it presents the first application of reinforcement learning to determine the mode of each slot. 
The results show that the amount of data transmitted in the network is affected by the HAP’s transmit power, the weight of downlink transmissions, distance between the HAP and each device and number of devices.
The reinforcement learning approach achieves $90\%$ of the optimal number of packet transmissions of the MILP. Moreover, it achieved $25\%$ more transmitted packets against benchmark approaches.
There are many possible future works.  The first is to apply and study advanced reinforcement learning methods to our problem.  The second is to develop more efficient methods to solve MINLP ($\mathbf{P1}$).

\bibliographystyle{IEEEtran}
\bibliography{Paper3.bib}

\begin{thebibliography}{10}
\providecommand{\url}[1]{#1}
\csname url@samestyle\endcsname
\providecommand{\newblock}{\relax}
\providecommand{\bibinfo}[2]{#2}
\providecommand{\BIBentrySTDinterwordspacing}{\spaceskip=0pt\relax}
\providecommand{\BIBentryALTinterwordstretchfactor}{4}
\providecommand{\BIBentryALTinterwordspacing}{\spaceskip=\fontdimen2\font plus
\BIBentryALTinterwordstretchfactor\fontdimen3\font minus \fontdimen4\font\relax}
\providecommand{\BIBforeignlanguage}[2]{{%
\expandafter\ifx\csname l@#1\endcsname\relax
\typeout{** WARNING: IEEEtran.bst: No hyphenation pattern has been}%
\typeout{** loaded for the language `#1'. Using the pattern for}%
\typeout{** the default language instead.}%
\else
\language=\csname l@#1\endcsname
\fi
#2}}
\providecommand{\BIBdecl}{\relax}
\BIBdecl

\bibitem{9141214}
S.~Niknam, H.~S. Dhillon, and J.~H. Reed, ``Federated learning for wireless communications: Motivation, opportunities, and challenges,'' \emph{IEEE Commun. Mag.}, vol.~58, no.~6, pp. 46--51, Jun. 2020.

\bibitem{9860057}
K.~Zhang, Y.~Shi, S.~Karnouskos, T.~Sauter, H.~Fang, and A.~W. Colombo, ``Advancements in industrial cyber-physical systems: An overview and perspectives,'' \emph{IEEE Trans. on Ind. Informa.}, vol.~19, no.~1, pp. 716--729, Jan. 2023.

\bibitem{powercast}
\BIBentryALTinterwordspacing
C.~Powercast, ``Wireless power products - powercastco.com,'' 2021. [Online]. Available: \url{https://www.powercastco.com}
\BIBentrySTDinterwordspacing

\bibitem{8214104}
T.~D. Ponnimbaduge~Perera, D.~N.~K. Jayakody, S.~K. Sharma, S.~Chatzinotas, and J.~Li, ``Simultaneous wireless information and power transfer ({SWIPT}): Recent advances and future challenges,'' \emph{IEEE Commun. Surveys Tuts}, vol.~20, no.~1, pp. 264--302, Dec. 2018.

\bibitem{9831440}
Y.~Mao, O.~Dizdar, B.~Clerckx, R.~Schober, P.~Popovski, and H.~V. Poor, ``Rate-splitting multiple access: Fundamentals, survey, and future research trends,'' \emph{IEEE Commun. Surveys Tuts.}, vol.~24, no.~4, pp. 2073--2126, 2022.

\bibitem{1421925}
J.~Andrews, ``Interference cancellation for cellular systems: a contemporary overview,'' \emph{IEEE Wireless Commun.}, vol.~12, no.~2, pp. 19--29, Apr. 2005.

\bibitem{485709}
B.~Rimoldi and R.~Urbanke, ``{A} rate-splitting approach to the {G}aussian multiple-access channel,'' \emph{IEEE Trans. Inf. Theory}, vol.~42, no.~2, pp. 364--375, Mar. 1996.

\bibitem{9528052}
Y.~Lu, K.~Xiong, P.~Fan, Z.~Zhong, B.~Ai, and K.~B. Letaief, ``Worst-case energy efficiency in secure {SWIPT} networks with rate-splitting {ID} and power-splitting {EH} receivers,'' \emph{IEEE Trans. Wireless Commun.}, vol.~21, no.~3, pp. 1870--1885, Sep. 2022.

\bibitem{9248582}
M.~R. Camana~Acosta, C.~E.~G. Moreta, and I.~Koo, ``Joint power allocation and power splitting for {MISO-RSMA} cognitive radio systems with {SWIPT} and information decoder users,'' \emph{IEEE Syst. J.}, vol.~15, no.~4, pp. 5289--5300, Nov. 2021.

\bibitem{s11276-019-02126-z}
M.~R. Camana, P.~V. Tuan, C.~E. Garcia, and I.~Koo, ``Joint power allocation and power splitting for {MISO} {SWIPT} {RSMA} systems with energy-constrained,'' \emph{Wireless Networks}, no.~26, pp. 2241--2254, Aug. 2020.

\bibitem{9627180}
T.~Li, H.~Zhang, X.~Zhou, and D.~Yuan, ``Full-duplex cooperative rate-splitting for multigroup multicast with {SWIPT},'' \emph{IEEE Trans. Wireless Commun.}, vol.~21, no.~6, pp. 4379--4393, Nov. 2022.

\bibitem{8491048}
A.~Salern and L.~Musavian, ``Rate splitting in multi-pair energy harvesting relaying systems,'' in \emph{15th ISWCS}, Lisbon, Portugal, Aug. 2018, pp. 1--5.

\bibitem{8815494}
Y.~Mao, B.~Clerckx, and V.~O. Li, ``Rate-splitting for multi-user multi-antenna wireless information and power transfer,'' in \emph{20th IEEE SPAWC}, Cannes, France, Aug. 2019, pp. 1--5.

\bibitem{9195473}
J.~Zhang, J.~Zhang, Y.~Zhou, H.~Ji, J.~Sun, and N.~Al-Dhahir, ``Energy and spectral efficiency tradeoff via rate splitting and common beamforming coordination in multicell networks,'' \emph{IEEE Trans. Commun.}, vol.~68, no.~12, pp. 7719--7731, Sep. 2020.

\bibitem{7555358}
H.~Joudeh and B.~Clerckx, ``Sum-rate maximization for linearly precoded downlink multiuser {MISO} systems with partial {CSIT}: {A} rate-splitting approach,'' \emph{IEEE Trans. Commun.}, vol.~64, no.~11, pp. 4847--4861, Aug. 2016.

\bibitem{9201435}
L.~Li, K.~Chai, J.~Li, and X.~Li, ``Resource allocation for multicarrier rate-splitting multiple access system,'' \emph{IEEE Access}, vol.~8, pp. 174\,222--174\,232, Sep. 2020.

\bibitem{9461768}
Z.~Yang, M.~Chen, W.~Saad, and M.~Shikh-Bahaei, ``Optimization of rate allocation and power control for rate splitting multiple access ({RSMA}),'' \emph{IEEE Trans. Commun.}, vol.~69, no.~9, pp. 5988--6002, Jun. 2021.

\bibitem{9663192}
A.~Mishra, Y.~Mao, O.~Dizdar, and B.~Clerckx, ``Rate-splitting multiple access for downlink multiuser {MIMO}: Precoder optimization and phy-layer design,'' \emph{IEEE Trans. Commun.}, vol.~70, no.~2, pp. 874--890, Dec. 2022.

\bibitem{9676684}
S.~A. Tegos, P.~D. Diamantoulakis, and G.~K. Karagiannidis, ``On the performance of uplink rate-splitting multiple access,'' \emph{IEEE Commun. Lett.}, vol.~26, no.~3, pp. 523--527, Jan. 2022.

\bibitem{9257190}
Z.~Yang, M.~Chen, W.~Saad, W.~Xu, and M.~Shikh-Bahaei, ``Sum-rate maximization of uplink rate splitting multiple access ({RSMA}) communication,'' \emph{IEEE Trans. Mobile Comput.}, vol.~21, no.~7, pp. 2596--2609, Nov. 2022.

\bibitem{9562976}
N.~Q. Hieu, D.~T. Hoang, D.~Niyato, and D.~I. Kim, ``Optimal power allocation for rate splitting communications with deep reinforcement learning,'' \emph{IEEE Wirel. Commun. Lett.}, vol.~10, no.~12, pp. 2820--2823, Oct. 2021.

\bibitem{7956255}
C.~Qin, W.~Ni, H.~Tian, R.~P. Liu, and Y.~J. Guo, ``Joint beamforming and user selection in multiuser collaborative {MIMO} {SWIPT} systems with nonnegligible circuit energy consumption,'' \emph{IEEE Trans. on Vehic. Tech.}, vol.~67, no.~5, pp. 3909--3923, Jun. 2018.

\bibitem{6884177}
C.~Xiong, L.~Lu, and G.~Y. Li, ``Energy efficiency tradeoff in downlink and uplink {TDD} {OFDMA} with simultaneous wireless information and power transfer,'' in \emph{IEEE ICC}, Sydney, NSW, Australia, Aug. 2014, pp. 5383--5388.

\bibitem{7833146}
P.~D. Diamantoulakis, K.~N. Pappi, G.~K. Karagiannidis, H.~Xing, and A.~Nallanathan, ``Joint downlink/uplink design for wireless powered networks with interference,'' \emph{IEEE Access}, vol.~5, pp. 1534--1547, Jan. 2017.

\bibitem{8854318}
J.~Tang, Y.~Yu, M.~Liu, D.~K.~C. So, X.~Zhang, Z.~Li, and K.-K. Wong, ``Joint power allocation and splitting control for {SWIPT}-enabled {NOMA} systems,'' \emph{IEEE Trans. on Wirel. Commun.}, vol.~19, no.~1, pp. 120--133, Oct. 2020.

\bibitem{7874074}
C.~Qin, W.~Ni, H.~Tian, and R.~P. Liu, ``Joint rate maximization of downlink and uplink in multiuser {MIMO} {SWIPT} systems,'' \emph{IEEE Access}, vol.~5, pp. 3750--3762, Mar. 2017.

\bibitem{8472716}
S.~Rezvani, N.~Mokari, and M.~R. Javan, ``Uplink throughput maximization in {OFDMA}-based {SWIPT} systems with data offloading,'' in \emph{ICEE}, Mashhad, Iran, Sep. 2018, pp. 572--578.

\bibitem{8294215}
Z.~Yang, W.~Xu, Y.~Pan, C.~Pan, and M.~Chen, ``Optimal fairness-aware time and power allocation in wireless powered communication networks,'' \emph{IEEE Trans. on Commun.}, vol.~66, no.~7, pp. 3122--3135, Feb. 2018.

\bibitem{8695878}
S.~Gong, S.~Ma, C.~Xing, and G.~Yang, ``Optimal beamforming and time allocation for partially wireless powered sensor networks with downlink {SWIPT},'' \emph{IEEE Trans. on Sig. Proc.}, vol.~67, no.~12, pp. 3197--3212, Apr. 2019.

\bibitem{S1874490721000616}
L.~Shufeng, W.~Zelin, and J.~Libiao, ``Joint rate maximization of downlink and uplink in {NOMA} {SWIPT} systems,'' \emph{Physical Communication}, vol.~46, no. 101324, Mar. 2021.

\bibitem{7982605}
K.~Lv, J.~Hu, Q.~Yu, and K.~Yang, ``Throughput maximization and fairness assurance in data and energy integrated communication networks,'' \emph{IEEE Internet Things J.}, vol.~5, no.~2, pp. 636--644, Jul. 2018.

\bibitem{9154671}
D.-T. Do, C.-B. Le, and F.~Afghah, ``Enabling full-duplex and energy harvesting in uplink and downlink of small-cell network relying on power domain based multiple access,'' \emph{IEEE Access}, vol.~8, pp. 142\,772--142\,784, Aug. 2020.

\bibitem{8855977}
M.~Syam, Y.~L. Che, S.~Luo, and K.~Wu, ``Joint downlink-uplink throughput optimization in wireless powered communication networks,'' in \emph{IEEE/CIC ICCC}, Changchun, China, Aug. 2019, pp. 852--857.

\bibitem{8947043}
------, ``Uplink throughput maximization for low latency in wireless powered communication networks,'' in \emph{IEEE 19th ICCT}, Xian, China, Oct. 2019, pp. 1002--1006.

\bibitem{8536389}
Z.~Na, M.~Zhang, M.~Jia, M.~Xiong, and Z.~Gao, ``Joint uplink and downlink resource allocation for the internet of things,'' \emph{IEEE Access}, vol.~7, pp. 15\,758--15\,766, Nov. 2019.

\bibitem{8292385}
B.~E. ElDiwany, A.~A. El-Sherif, and T.~ElBatt, ``Optimal uplink and downlink resource allocation for wireless powered cellular networks,'' in \emph{IEEE 28th PIMRC}, Montreal, Canada, Oct. 2017, pp. 1--6.

\bibitem{8650384}
J.~Kim, H.~Lee, S.-H. Park, and I.~Lee, ``Joint downlink and uplink design for wireless powered cloud radio access networks,'' in \emph{IEEE TENCON}, Jeju, South Korea, Oct. 2018, pp. 1313--1316.

\bibitem{S1389128620311476}
W.~Mohammed, A.~Emad, and S.~Yao, ``Resource allocation for {SWIPT}-enabled energy-harvesting downlink/uplink clustered {NOMA} networks,'' \emph{Computer Networks}, vol. 182, no. 107471, Dec. 2020.

\bibitem{7848950}
Z.~Liang and Y.~Liu, ``Link scheduling in {SWIPT} systems,'' in \emph{IEEE GC Wkshps}, Washington, DC, USA, Dec. 2016, pp. 1--6.

\bibitem{7990524}
Q.~Yao, T.~Q.~S. Quek, A.~Huang, and H.~Shan, ``Joint downlink and uplink energy minimization in wet-enabled networks,'' \emph{IEEE Trans. Wireless Commun.}, vol.~16, no.~10, pp. 6751--6765, Jul. 2017.

\bibitem{10109156}
X.~Song and K.-W. Chin, ``Maximizing packets collection in wireless powered {IoT} networks with charge-or-data time slots,'' \emph{IEEE Trans. Cogn. Commun. Netw.}, pp. 1--1, Apr. 2023.

\bibitem{9367293}
------, ``A novel hybrid access point channel access method for wireless-powered {IoT} networks,'' \emph{IEEE Internet Things J.}, vol.~8, no.~15, pp. 12\,329--12\,338, Mar. 2021.

\bibitem{8388904}
B.~Liu, Y.~Wei, L.~Liu, R.~Hu, and G.~Qiao, ``A joint uplink/downlink resource allocation algorithm in {OFDMA} wireless networks,'' in \emph{9th ICAIT}, Chengdu, China, Nov. 2017, pp. 143--149.

\bibitem{6678102}
H.~Ju and R.~Zhang, ``Throughput maximization in wireless powered communication networks,'' \emph{IEEE Trans. Wireless Commun.}, vol.~13, no.~1, pp. 418--428, Dec. 2014.

\bibitem{10198464}
G.~Zheng, M.~Wen, Y.~Chen, Y.-C. Wu, and H.~Vincent~Poor, ``Rate-splitting multiple access in wireless backhaul {HetNets}: A decentralized spectral efficient approach,'' \emph{IEEE Trans. on Wirel. Commun.}, vol.~23, no.~3, pp. 2413--2427, 2023.

\bibitem{6489506}
R.~Zhang and C.~K. Ho, ``{MIMO} broadcasting for simultaneous wireless information and power transfer,'' \emph{IEEE Trans. Wireless Commun.}, vol.~12, no.~5, pp. 1989--2001, Mar. 2013.

\bibitem{7264986}
E.~Boshkovska, D.~W.~K. Ng, N.~Zlatanov, and R.~Schober, ``Practical non-linear energy harvesting model and resource allocation for {SWIPT} systems,'' \emph{IEEE Commun. Lett.}, vol.~19, no.~12, pp. 2082--2085, Sep. 2015.

\bibitem{8714026}
N.~C. Luong, D.~T. Hoang, S.~Gong, D.~Niyato, P.~Wang, Y.-C. Liang, and D.~I. Kim, ``Applications of deep reinforcement learning in communications and networking: A survey,'' \emph{IEEE Commun. Surveys Tuts.}, vol.~21, no.~4, pp. 3133--3174, May 2019.

\bibitem{5522465}
S.~Sudevalayam and P.~Kulkarni, ``Energy harvesting sensor nodes: Survey and implications,'' \emph{IEEE Commun. Surveys Tuts.}, vol.~13, no.~3, pp. 443--461, Jul. 2011.

\bibitem{4460126}
J.-S. Lee, Y.-W. Su, and C.-C. Shen, ``A comparative study of wireless protocols: Bluetooth, {UWB}, {ZigBee}, and {Wi-Fi},'' in \emph{33rd Annual Conference of the IEEE Industrial Electronics Society}, Taipei, Taiwan, Nov. 2007, pp. 46--51.

\end{thebibliography}

\end{document}